\documentclass[conference]{IEEEtran}
\usepackage{mathtools}
\usepackage{cite}
\usepackage{xcolor}
\usepackage{caption}
\usepackage{subcaption}
\usepackage{amsmath, amssymb}
\usepackage{amsthm}
\usepackage{pifont}
\usepackage{booktabs}
\usepackage[utf8]{inputenc}
\usepackage{algorithm}
\usepackage{algpseudocode}
\usepackage{amsfonts}
\usepackage{graphicx}
\usepackage{xcolor}
\usepackage{bbm}
\usepackage{listings}
\usepackage{bm}
\usepackage{multirow}
\usepackage{lipsum}
\usepackage{comment} 

\newtheorem{theorem}{Theorem}
\title{Joint User Pairing and Association for Multicell NOMA: A Pointer Network-based Approach}
\author{\IEEEauthorblockN{Manyou Ma and Vincent W.S. Wong}
\IEEEauthorblockA{Department of Electrical and Computer Engineering, The University of British Columbia, Vancouver, Canada}
email: \{manyoum, vincentw\}@ece.ubc.ca
}

\begin{document}
\maketitle
\begin{abstract}
In this paper, we investigate the joint user pairing and association problem for multicell non-orthogonal multiple access (NOMA) systems. We consider a scenario where the user equipments (UEs) are located in a multicell network equipped with multiple base stations. Each base station has multiple orthogonal physical resource blocks (PRBs). Each PRB can be allocated to a pair of UEs using NOMA. Each UE has the additional freedom to be served by any one of the base stations, which further increases the complexity of the joint user pairing and association algorithm design. Leveraging the recent success on using machine learning to solve numerical optimization problems, we formulate the joint user pairing and association problem as a combinatorial optimization problem. The solution is found using an emerging deep learning architecture called Pointer Network (PtrNet), which has a lower computational complexity compared to solutions based on iterative algorithms and has been proven to achieve near-optimal performance. The training phase of the PtrNet is based on deep reinforcement learning (DRL), and does not require the use of the optimal solution of the formulated problem as training labels. Simulation results show that the proposed joint user pairing and association scheme achieves near-optimal performance in terms of the aggregate data rate, and outperforms the random user pairing and association heuristic by up to 30\%.
\end{abstract}

\section{Introduction}
% Motivation
Non-orthogonal multiple access (NOMA) has been proposed as an enabling technology for the fifth generation (5G) wireless networks~\cite{wong2017key}. Compared to the conventional orthogonal multiple access (OMA) approach, where each user equipment (UE) is allocated a single physical resource block (PRB), NOMA allows multiple UEs to share one PRB. NOMA has shown promise in improving the UE connection density, spectral efficiency, and user fairness in wireless networks~\cite{Shin2018Non}.  Downlink power-domain NOMA is a genre of NOMA realization, where a base station can transmit data packets to multiple UEs by superimposing their signals in the power domain. Algorithms such as successive interference cancellation (SIC) can be used by the UE to decode the signal. One of the key issues of power-domain NOMA is to find the optimal \textit{user pairing} policy, which determines the mapping between a pair of UEs and a particular PRB. Moreover, multicell NOMA has also been proposed~\cite{Shin2018Non, wang2019user, baidas2019user,you2018resource}, where the paired UEs have the freedom to connect to one of the base stations in the network. Therefore, the \textit{user association} policy, which decides a UE to be served by which particular base station, also needs to be studied.

Various user pairing strategies have been proposed in the literature. In~\cite{Ding2016impact}, Ding \textit{et al.} proposed a user pairing strategy based on sorting the channel state information (CSI) of the UEs.  Optimization-based methods, such as monotonic optimization~\cite{sun2017optimal} and difference-of-convex programming~\cite{mostafa2019connection}, have been used to find the optimal solution of the user pairing problem in NOMA. Due to the nonconvex nature of user pairing problem, this type of problem is computationally intensive to solve and the near-optimal solution is only attainable for problems with special structure, (\textit{e.g.,} monotonicity or the ability to be expressed in a difference-of-convex form). Joint user pairing and association optimization algorithms designed for multicell NOMA have also been studied. Some of these algorithms~\cite{wang2019user, baidas2019user} are iterative and converge to the optimal solution only after a large number of iterations. 

Recently, machine learning and deep learning (DL)-based architectures have begun to be adopted to solve numerical optimization problems, and have been proven to have lower computational complexity than some of the existing approaches~\cite{Vinyals2015Pointer}. Recurrent neural network-based architectures such as sequence-to-sequence (\textit{seq-2-seq}) model~\cite{sutskever2014sequence} and the attention mechanism~\cite{bahdanau2014neural}, which are originally proposed for natural language processing applications, are beginning to be adopted to solve classical combinatorial optimization problems~\cite{Vinyals2015Pointer, bello2017neural}. In~\cite{bello2017neural}, it has been shown that combined with deep reinforcement learning (DRL), the trained deep neural networks can achieve near-optimal performance in solving the traveling salesman and knapsack problems. 

Leveraging the recent success on using DL to solve combinatorial optimization problems, in this work, we propose a joint user pairing and association algorithm for multicell NOMA using a DL-based approach. Specifically, we use a network structure called \textit{pointer network} (PtrNet)~\cite{Vinyals2015Pointer, bello2017neural}. Compared to the exhaustive search~\cite{zhu2019optimal} and iterative game-theoretic approaches~\cite{baidas2019user, wang2019user} in the literature, the proposed PtrNet-based solution is non-iterative, has a lower computational complexity, and achieves near-optimal performance. The contributions of our work are as follows:
\begin{itemize}
    \item We formulate the joint user pairing and association problem for NOMA as a combinatorial optimization problem. The input (or parameter) of the optimization problem corresponds to the CSI between a UE and a base station in the network. The solution to the optimization problem corresponds to the joint user pairing and association decisions for all the UEs. 
    
    \item We propose using the state-of-the-art PtrNet~\cite{Vinyals2015Pointer} architecture to solve the combinatorial optimization problem, and use a REINFORCE-based method~\cite{bello2017neural} to perform parameter optimization of the network. The optimal solutions are not required for the training of the PtrNet.
    
    \item Our simulation results show that the proposed PtrNet-based solution achieves near-optimal performance, with an optimality gap less than 2\% compared to the optimal joint user pairing and association strategy obtained from exhaustive search. The proposed PtrNet-based solution outperforms the random pairing NOMA heuristic by 30\% in terms of the aggregate data rate.

\end{itemize}

The rest of this paper is organized as follows. In Section II, we introduce the system model and present the data rate expressions. In Section III, we formulate the joint user pairing and association problem as a combinatorial optimization problem and transform it into a permutation-finding problem. In Section IV, we present a PtrNet-based solution to solve the formulated problem. Simulation results are presented in Section V. Section VI concludes the paper. 

% ----------------------------------------------------------------------------------------
\section{System Model}
Consider a multicell network with $N$ UEs and $K$ base stations. Each base station is allocated with $B$ PRBs. We consider the case where $N = 2BK$. If the number of active users in the network is greater than $2BK$, then the admission control module in the network will accept at most $2BK$ users\footnote{When the number of active users in the network is less than $2BK$, then we introduce surrogate users with zero CSI to all base stations to ensure that $N=2BK$.}. Let $\mathcal{N} = \{1, \ldots, N\}$ denote the set of UEs and $\mathcal{K} = \{1, \ldots, K\}$ denote the set of base stations. Different base stations are allocated with orthogonal sets of PRBs and an efficient frequency reuse scheme is adopted so that inter-cell interference becomes negligible. Let set  $\mathcal{B}_k = \{1, \ldots, B\}$ denote the PRBs allocated to base station $k\in\mathcal{K}$. We consider a time-slotted system, where the user pairing and association decision is made at the beginning of each time slot. 

\begin{figure}[t]
    \centering
    \includegraphics[width=0.45\textwidth]{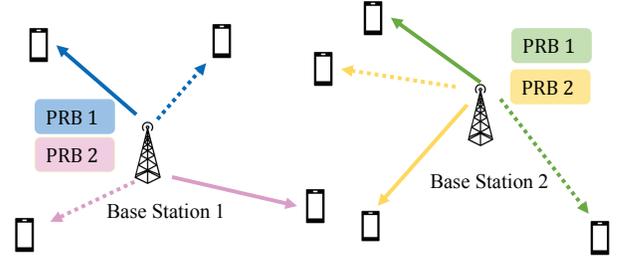}
        \caption{Illustration of a sample multicell network with two base stations and eight UEs. Each base station is allocated with two PRBs. NOMA is employed which enables each PRB to serve two UEs, one as an SIC user (denoted by a dashed arrow) and the other one as a non-SIC user (denoted by a solid arrow).}
    \label{fig:areadiagram}
    \vspace{-0.5cm}
\end{figure}

When NOMA is employed, each PRB can serve two UEs in total. One UE is served as an \textit{SIC user} and the other UE is served as a \textit{non-SIC user}. The non-SIC user directly decodes its own signal while treating the signal intended for the SIC user as interference. On the other hand, the SIC user first removes the signal intended for the non-SIC user before decoding its own signal. With $K$ based stations, each having $B$ PRBs, the network can serve $2BK$ UEs simultaneously using NOMA. An illustration of the network topology is shown in Fig.~\ref{fig:areadiagram}.

Let $|h_{k, n}|^2 \in \mathbb{R}_+$ denote the CSI between base station $k\in \mathcal{K}$ and UE $n\in \mathcal{N}$. We consider a narrowband scenario, where the CSI between base station $k$ and UE $n$ does not depend on the allocated PRB. Moreover, let $\mathbf{h}_n = (|h_{1, n}|^2, \ldots, |h_{K, n}|^2) \in \mathcal{H} \subseteq \mathbb{R}^{K}$ denote the CSI vector of UE $n$. Let $\mathbf{H} \in \mathbb{R}^{K \times N}$ denote the CSI matrix of the system. We have
\begin{equation}
\mathbf{H} = [\mathbf{h}_1, \ldots, \mathbf{h}_N] \in \mathcal{H}^N \subseteq \mathbb{R}^{K\times N}.
\end{equation}
We consider large-scale fading with pathloss exponent $\beta$ and small-scale Rayleigh fading. Let $l_{k,n}$ denote the distance between base station $k$ and UE $n$. Given $l_{k,n}$, we have $|h_{k, n}|^2 = l_{k,n}^{-\beta}\mathcal{E}$, where $\mathcal{E}$ is drawn from an exponential distribution. The locations of the base stations are known, and the locations of the UEs are drawn from a random distribution.

\subsection{Joint User Pairing and Association Matrix}
Let us define a binary joint user pairing association matrix $\mathbf{X} \in \{0,1\}^{K\times N \times B \times 2}$, where $x_{k, n, b, 1} = 1$ if UE $n$ is associated with base station $k$ and is allocated with PRB $b$ as an SIC user, and is equal to zero otherwise. Similarly, $x_{k, n, b, 2} = 1$ if UE $n$ is associated with base station $k$ and is allocated with PRB $b$ as a non-SIC user, and is equal to zero otherwise. Since $\mathbf{X}$ is a binary matrix, we need 
\begin{equation}
x_{k,n, b, p} \in \{0,1\}, \quad  n\in\mathcal{N},\ b \in \mathcal{B}_k,\  k\in\mathcal{K},\ p\in\{1,2\}.
\label{eq:binary_constraint}
\end{equation}
Since each UE $n$ can only be associated with one base station and be allocated with one PRB, either as an SIC or a non-SIC user, we have 
\begin{equation}
\sum_{k=1}^{K}\sum_{b=1}^{B}\sum_{p=1}^{2}x_{k,n,b,p} = 1,  \quad   n\in\mathcal{N}.
\label{eq:one_decision_constraint}
\end{equation}
Since each PRB $b$ in base station $n$ can only serve one SIC user and one non-SIC user, we have
\begin{equation}
\sum_{n=1}^{N} x_{k,n,b,p} = 1, \quad   b\in \mathcal{B}_k,\ k\in\mathcal{K},  \ p\in\{1,2\}.
\label{eq:one_user_constraint}
\end{equation}
\subsection{Data Rate Expression}
Suppose UE $n\in \mathcal{N}$ is associated with base station $k\in\mathcal{K}$, the maximum achievable data rate using OMA is
\begin{equation}
R^\text{OMA}_{k, n} = \frac{1}{2}\log_2\left(1+{\frac{P|h_{k,n}|^2}{\sigma^2}}\right), 
    \label{eq:OMA_SINR}
\end{equation}
\noindent where $P$ denotes the transmit power of the base station and $\sigma^2$ denotes the noise variance at the receiver. The $\frac{1}{2}$ factor in~(\ref{eq:OMA_SINR}) results from the $\frac{1}{2}$ multiplexing loss of OMA, compared to NOMA.

We define a conservative minimum rate constraint for UE $n$ when NOMA is used, $\gamma_n$, as the minimum achievable data rate using OMA when associated with any of the base stations. That is, given UE $n\in\mathcal{N}$, its minimum rate requirement is
\begin{equation}
\gamma_{n} = \min\left(R^\text{OMA}_{1, n}, \ldots, R^\text{OMA}_{K, n}\right).
    \label{eq:OMA_guaranteedR}
\end{equation}

Let $\mathbf{A} \in [0,1]^{K \times B}$ denote the matrix of power allocation coefficients, where $\alpha_{k,b} \in [0,1]$ denotes the power allocation coefficient for the SIC user associated with base station $k\in\mathcal{K}$ and allocated with PRB $b \in \mathcal{B}_k$. Given $\alpha_{k,b}$ and that UE $n$ is associated with base station $k$ and is allocated with PRB $b$ as a non-SIC user, the maximum data rate it can achieve is as follows:
\begin{equation}
R_{k, n, b, 2} = \log_2\left(1+ \frac{(1-\alpha_{k,b})  |h_{k,n}|^2 P}{\alpha_{k,b}P|h_{k,n}|^2+\sigma^2} \right).
\label{eq:NOMA_SINR_nSIC}
\end{equation}
Similarly, given $\alpha_{k,b}$ and that UE $n$ is associated with base station $k$ as an SIC user and is allocated with PRB $b$, the maximum data rate it can achieve is as follows:
\begin{equation}
R_{k, n, b, 1} = \log_2\left(1+ \frac{\alpha_{k,b}P|h_{k,n}|^2}{\sigma^2} \right).
\label{eq:NOMA_SINR_SIC}
\end{equation}
For SIC user $n$ to achieve the data rate expressed in (\ref{eq:NOMA_SINR_SIC}), it needs to first successfully decode the signal intended for the non-SIC user allocated with the same PRB. That is,
\begin{equation}
\sum_{n=1}^{N} x_{k,n,b,1}|h_{k, n}|^2 \geq \sum_{n=1}^{N} x_{k,n,b,2}|h_{k, n}|^2, b\in\mathcal{B}_k, k\in\mathcal{K}.
\label{eq:SIC_constraint}
\end{equation}
\section{Problem Formulation}
\subsection{Joint User Pairing and Association Problem}
The joint user pairing and association problem can be formulated as the following data rate maximization problem:
\begin{subequations}
\begin{alignat}{2}
&\underset{\{\mathbf{X}, \bm{A} \}}{\text{maximize} }       & & \Phi(\mathbf{X}, \bm{A}\ |\ \mathbf{H})
= \sum_{k=1}^{K}\sum_{n=1}^{N}\sum_{b=1}^{B}\sum_{p=1}^{2}x_{k,n,b,p}R_{k, n, b, p},\\
&\text{subject to} & \quad & 0 \leq \alpha_{k,b} \leq 1, \quad  b \in \mathcal{B}_k, k \in \mathcal{K}\\
& & \quad & \sum_{k=1}^{K}\sum_{b=1}^{B}\sum_{p=1}^{2}x_{k,n, b, p}R_{k, n, b, p} \geq \gamma_n,  n\in\mathcal{N}\label{eq:OMA_rate_constraint}\\
& & \quad & \text{constraints (\ref{eq:binary_constraint}), (\ref{eq:one_decision_constraint}), (\ref{eq:one_user_constraint}), (\ref{eq:SIC_constraint})},\nonumber\end{alignat}
\label{eq:optimiztion_NOMA}
\end{subequations}
where $\Phi(\mathbf{X}, \mathbf{A}\ |\ \mathbf{H})$ is the aggregate data rate of the network, given the CSI matrix $\mathbf{H}$. Constraint (\ref{eq:OMA_rate_constraint}) is chosen to be the minimum rate constraints for all UEs. We point out that problem (\ref{eq:optimiztion_NOMA}) is a nonconvex mixed-integer programming problem, where the optimal solution is hard to find. In the next subsection, we will transform the problem into a pure combinatorial optimization problem.
\subsection{Transforming Problem (\ref{eq:optimiztion_NOMA}) into a Combinatorial Problem} 
Let us consider decomposing problem (\ref{eq:optimiztion_NOMA}) into two sub-problems. The first sub-problem is to find the joint user pairing and association decision matrix $\mathbf{X}$. The second sub-problem is to find the optimal power allocation coefficients matrix $\mathbf{A}$. Given the solution of the first sub-problem, where $n$ and $m \in \mathcal{N}$ denote the SIC and non-SIC users that are associated with base station $k$ and are allocated with PRB $b$ such that $|h_{k,n}|^2 \geq |h_{k,m}|^2$, respectively, then the optimal power allocation coefficient $\alpha^*_{k,b}$ can be found by solving the following optimization problem:
\begin{subequations}
\begin{alignat}{2}
R_{k,b}(m,n) = & \underset{ \alpha_{k,b}}{\ \text{maximize} }       & &R_{k,n,b,1} + R_{k,m,b,2}\\
&\ \text{subject to} & \quad & 0 \leq \alpha_{k,b} \leq 1 \\
& & \quad & R_{k, n, b, 1} \geq \gamma_n
\label{eq:power_allocation_n_constraint}\\
& & \quad & R_{k, m, b, 2} \geq \gamma_m.
\label{eq:power_allocation_m_constraint}
\end{alignat}
\label{eq:powerAllocation}
\end{subequations}
Following the analysis in \cite{zhu2019optimal}, the optimal power allocation coefficient in problem (\ref{eq:powerAllocation}) has a closed-form solution, given by the following theorem. 
\begin{theorem}
For paired NOMA users $n$ and $m\in \mathcal{N}$ are associated with base station $k \in \mathcal{K}$ and are allocated with PRB $b \in \mathcal{B}_k$, such that $|h_{k,n}|^2 \geq |h_{k,m}|^2$, the optimal power allocation coefficient, subject to the minimum rate constraint, can be expressed as
\begin{equation}
\alpha_{k,b}^*(m, n) = \frac{(1+|h_{k, m}|^2\eta)/\sqrt{1+|h_{j, m}|^2\eta}-1}{|h_{k,m}|^2\eta},
\label{eq:optimal_power_coeff}
\end{equation}
\noindent where $\eta = \frac{P}{\sigma^2}$ denotes the signal-to-noise ratio (SNR) and $j \in \mathcal{K}$ satisfies $R^\text{OMA}_{j,m} = \gamma_m$.
\end{theorem}
\begin{proofsketch} In~\cite{zhu2019optimal}, it was proven that the objective function is nondecreasing with respect to $\alpha_{k, b}$. The lower bound of $\alpha_{k,b}^*(m, n)$ is due to constraint (\ref{eq:power_allocation_n_constraint}), while the upper bound is due to constraint (\ref{eq:power_allocation_m_constraint}). The optimal solution is the upper bound, that is
\begin{align*}
& R_{k, m, b, 2} \geq \gamma_m = R^\text{OMA}_{j,m}\\
\Leftrightarrow\ & \log_2\left(1+ \frac{(1-\alpha_{k,b})  |h_{k,m}|^2 }{\alpha_{k,b}|h_{k,m}|^2+\eta^{-1}} \right) \geq \frac{1}{2}\log_2\left(1+|h_{j,m}|^2\eta\right)\\
\Leftrightarrow\ & \alpha_{k,b}(m, n) \leq \frac{(1+|h_{k, m}|^2\eta)/\sqrt{1+|h_{j, m}|^2\eta}-1}{|h_{k,m}|^2\eta}.
\end{align*}
We can show constraint (\ref{eq:power_allocation_n_constraint}) is satisfied based on Theorem 1 in \cite{zhu2019optimal}. \quad \textbf{Q.E.D.}
\end{proofsketch}

To simplify the notations, we use a matrix $\mathbf{M} \in \mathcal{N}^{K\times B\times 2}$ to denote the joint user pairing and association decision. We use $m_{k,b,1}$ and $m_{k, b, 2}$ to denote the SIC and non-SIC users that are associated with base station $k$ and are allocated with PRB $b$, respectively. That is 
\begin{equation*}
m_{k,b,p} \in \big\{n\ |\ n\in\mathcal{N}, x_{k, n, b, p} = 1\big\},  b\in\mathcal{B}_k, k\in \mathcal{K}, p\in \{1,2\}.
\end{equation*}
Since each $\mathbf{X}$ corresponds to a unique $\mathbf{M}$, we will use them interchangeably to denote the joint user pairing and association decision. By using the optimal power allocation coefficients, problem (\ref{eq:optimiztion_NOMA}) is reduced to the following combinatorial optimization problem:
\begin{subequations}
\begin{alignat}{2}
&\underset{\mathbf{X}}{\text{maximize} }       & & \Phi(\mathbf{X}\ |\ \mathbf{H}) = \sum_{k=1}^{K}\sum_{b=1}^{B} R_{k,b}(m_{k,b,2}, m_{k,b,1} )\\
&\text{subject to} & \quad &  \alpha_{k,b} = \alpha_{k,b}^*(m_{k,b,2}, m_{k,b,1}),  b \in \mathcal{B}_k, k \in \mathcal{K}\\
& & \quad & \text{constraints (\ref{eq:binary_constraint}), (\ref{eq:one_decision_constraint}), (\ref{eq:one_user_constraint}), (\ref{eq:SIC_constraint})}.\nonumber
\end{alignat}
\label{eq:optimiztion_combinatorial}
\end{subequations}
\noindent 
Although problem (\ref{eq:optimiztion_combinatorial}) is a pure combinatorial optimization problem, its optimal solution is difficult to obtain. Previous research has considered exhaustive search~\cite{zhu2019optimal} and game-theoretic approach~\cite{wang2019user, baidas2019user}. However, these methods are both iterative and may incur a high computational complexity.

Finally, we point out that problem (\ref{eq:optimiztion_combinatorial}) is equivalent to the following optimization problem:
\begin{subequations}
\begin{alignat}{2}
&\underset{\mathbf{X}}{\text{maximize} }       & & \Phi(\mathbf{X}\ |\ \mathbf{H}) = \sum_{k=1}^{K}\sum_{b=1}^{B} R_{k,b}(m_{k,b}^{-}, m_{k,b}^{+})\\
&\text{subject to} & \quad &  \alpha_{k,b} = \alpha_{k,b}^*(m_{k,b}^{-}, m_{k,b}^{+}),  b \in \mathcal{B}_k, k \in \mathcal{K}\\
& & \quad & m_{k,b}^{+} = \underset{i \in \{m_{k,b,1}, m_{k,b,2}\}}{\arg\max}\ |h_{k,i}|^2, b \in \mathcal{B}_k, k\in \mathcal{K}\nonumber\\
& & \quad & m_{k,b}^{-} = \underset{i \in \{m_{k,b,1},  m_{k,b,2}\}}{\arg\min}\ |h_{k,i}|^2, b \in \mathcal{B}_k, k\in \mathcal{K}\nonumber\\
& & \quad & \text{constraints (\ref{eq:binary_constraint}), (\ref{eq:one_decision_constraint}), (\ref{eq:one_user_constraint})}.\nonumber
\end{alignat}
\label{eq:optimiztion_permutation}
\end{subequations}
\noindent In problem (\ref{eq:optimiztion_permutation}), the UE with better CSI within a pair is chosen to be the SIC user. In this case, constraint (\ref{eq:SIC_constraint}) can be relaxed. 
Constraints (\ref{eq:binary_constraint}), (\ref{eq:one_decision_constraint}), and  (\ref{eq:one_user_constraint}) specify that each element in set $\mathcal{N}$ appears in matrix $\mathbf{M}$ only once. Since there are $N= 2BK$ entries in $\mathbf{M}$, the necessary and sufficient condition for constraints (\ref{eq:binary_constraint}), (\ref{eq:one_decision_constraint}) and  (\ref{eq:one_user_constraint}) to be satisfied is $\mathbf{M}$ being a permutation of elements in $\mathcal{N}$.  One way to obtain the optimal $\mathbf{M}$ is to exhaustively search through all the permutations of all elements in set $\mathcal{N}$. However, this quickly becomes computationally intractable as $N$ increases. In the next section, we propose to use the \textit{seq-2-seq} framework to find the optimal permutation of elements from set $\mathcal{N}$, with a lower computational complexity.

\section{Solution Using the PtrNet Architecture}
In this section, we solve problem (\ref{eq:optimiztion_permutation}) using the \textit{seq-2-seq} framework and the PtrNet architecture. We start with the problem setup and then present the PtrNet architecture and the REINFORCE-based training method. 

\subsection{Finding the Optimal Permutation Using \textit{seq-2-seq}}
Let us define the input sequence as
\begin{equation}
\mathbf{s} = (\mathbf{h}_1, \mathbf{h}_2, \ldots, \mathbf{h}_N) \in \mathcal{H}^{N},
\end{equation}
\noindent and the output sequence as
\begin{equation}
\begin{split}
\mathbf{u} = &\ (m_{1, 1, 1}, m_{1, 1, 2} \ldots, m_{1, B, 1}, m_{1, B, 2},  \\
&\quad \ldots, m_{K, B, 1}, m_{K, B, 2}) = (u_1, \ldots, u_{N})  \in \mathcal{N}^{N}, 
\end{split}
\label{eq:NOMA_action_vec}
\end{equation} 
\noindent which is a permutation of the elements in set $\mathcal{N}$. That is, 
\begin{equation}
u_{i} \neq u_{j},  \forall i\neq j, \quad i, j \in \mathcal{N}.
\end{equation}
Since each sequence $\mathbf{s}$ corresponds to  a CSI matrix $\mathbf{H}$ and each sequence $\mathbf{u}$ corresponds to a joint user pairing and association matrix $\mathbf{X}$, we will use them ($\mathbf{s}$ 
\textit{v.s.} $\mathbf{H}$ and $\mathbf{u}$ \textit{v.s.} $\mathbf{X}$) interchangeably in the remainder of this paper. Hence, the aggregate data rate of an input sequence $\mathbf{s}$ and the output sequence $\mathbf{u}$ pair can be expressed as 
$\Phi(\mathbf{u}\ |\ \mathbf{s} )$.

Using the \textit{seq-to-seq} model, we aim to find a parametric model, which is a PtrNet with parameters $\bm{\theta}$, that computes the conditional probability $p_{\bm{\theta}}(\mathbf{u}\ |\ \mathbf{s})$, for each pair of $\mathbf{u}$ and $\mathbf{s}$, by using the \textit{probability chain rule}
\begin{equation}
p_{\bm{\theta}}(\mathbf{u}\ |\ \mathbf{s}) = \prod_{n=1}^N p_{ \bm{\theta}}(u_n\ |\ u_1, \ldots, u_{n-1}, \mathbf{s}). 
\end{equation}
During the training phase, the optimal parameters $\bm{\theta}^*$  are learned to optimize the expected aggregate data rate, such that
\begin{equation}
\bm{\theta}^* = \underset{\bm\theta}{\arg \max}\bigg\{\underset{\mathbf{u}\sim p_{\bm{\theta}(\cdot|\mathbf{s}),\mathbf{s}\sim\mathcal{L}}}{\mathbb{E}}\left[\Phi(\mathbf{u}\ |\ \mathbf{s})\right]\bigg\}, 
\end{equation}
\noindent where $\mathbb{E}[\cdot]$ denotes the expectation of a random variable and $\mathcal{L}$ is the probability distribution of the input sequence $\mathbf{s}$. After training, given an input sequence $\mathbf{s}$, the output sequence $\mathbf{u}$ can be sampled one element at a time, based on the policy parameterized by $\bm{\theta}^*$, where
\begin{equation}
u_n \sim  p_{\bm{\theta}^*}(u_n\ |\ u_1, \ldots, {u}_{n-1}, \mathbf{s}) \quad n\in\mathcal{N}.
\end{equation}
\subsection{Overview of the PtrNet}
\begin{figure}[t]
    \centering
    \includegraphics[width=0.5\textwidth]{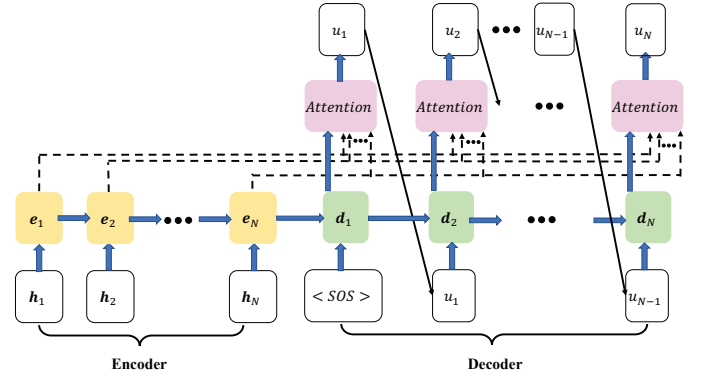}
        \caption{Structure of the pointer network (PtrNet).}
    \label{fig:overview_ptrNet}
    \vspace{-0.5cm}
\end{figure}

In this subsection, we introduce the PtrNet architecture, which we chose to parameterize $p_{\bm{\theta}}(\mathbf{u}\ |\ \mathbf{s})$. We first introduce its components: the \textit{encoder}, the \textit{decoder}, and the \textit{attention} module, as illustrated in Fig.~\ref{fig:overview_ptrNet}.

\subsubsection{Encoder and Decoder}
Following the notations from the \textit{seq-2-seq} model~\cite{sutskever2014sequence}, we adopt a two-stage approach to generate the output sequence: an encoding stage and a decoding  stage. We train two separate neural networks to achieve this: an encoder for the encoding stage and a decoder for the decoding stage. Both the encoder and decoder are implemented using long short-term memory (LSTM) cells. 

Given an input sequence with length $N$, the encoder takes elements from the input sequence one element at a time, and generates a sequence of hidden states,
\begin{equation}
(\mathbf{e}_1, \mathbf{e}_2, \ldots, \mathbf{e}_N), 
\end{equation}
\noindent in $N$ encoding steps, where $\mathbf{e}_n \in \mathbb{R}^H$ and $H$ denotes the dimension of the hidden units of the LSTM cells. Let the encoder parameters be denoted by $\bm{\theta}_e$. Then, at the encoding step $n \in \mathcal{N}$, the hidden state $\mathbf{e}_n$ depends on the current input and the hidden state from the previous encoding step. That is,
\begin{equation}
\mathbf{e}_n = f_{\bm{\theta}_e}(\mathbf{e}_{n-1}, \mathbf{h}_n), \ \ \  n \in \mathcal{N},
\end{equation}
\noindent where $f_{\bm{\theta}_e}$ is a function parameterized by parameters $\bm{\theta}_e$. We define $\mathbf{e}_0$ as the initial state of the encoder. 

The decoder generates a hidden sequence of length $N$, 
\begin{equation}
(\mathbf{d}_1, \mathbf{d}_2, \ldots, \mathbf{d}_{N}),
\end{equation}
one element at a time in $N$ decoding steps, where $\mathbf{d}_n \in \mathbb{R}^H$, $ n\in \mathcal{N}$. We define a special symbol $\langle SOS \rangle$, which specifies the start of the output sequence.

Let the decoder parameters be denoted by $\bm{\theta}_d$. Then, at each decoding step $n$, the decoder generates the hidden state $\mathbf{d}_n$, based on the last output of the decoder $\mathbf{u}_{n-1}$ and the hidden state from the last decoding step $\mathbf{d}_{n-1}$. That is
\begin{equation}
\mathbf{d}_n = f_{\bm{\theta}_d}(\mathbf{d}_{n-1}, \mathbf{u}_{n-1}), \ \ \  n \in \mathcal{N},
\end{equation}
where $f_{\bm{\theta}_d}$ is a function parameterized by parameters $\bm{\theta}_d$. We define $\mathbf{d}_0 = \mathbf{e}_N$ and ${u}_0 = \langle SOS \rangle$. 

\subsubsection{Pointing Mechanism}
For the joint user pairing and association problem,  the output sequence corresponds to indices of the input sequence. Therefore, at each decoding step, we use the \textit{attention mechanism}~\cite{bahdanau2014neural} to determine the probability of pointing to one of the elements from set  $\mathcal{N}$ as the output.

At decoding step $n$, the first $n-1$ entries of the output sequence $\mathbf{u}_{n-1} = (u_1, u_2, \ldots, u_{n-1})$ have already been generated. Let set $\mathcal{U}^{n-1}$ include all the $n-1$ entries in $\mathbf{u}_{n-1}$. We first calculate the \textit{attention mask}, which ensures that there are no repeated elements in $\mathbf{u}$, based on the previous output $\mathbf{u}_{n-1}$. Let $\mathbf{Q}^n = (q^n_1, q^n_2, \ldots, q^n_N)$ denote the attention mask, where $q^n_j \in \{0,1\}$, $j \in \mathcal{N}$. We have 
\begin{equation}
    q_j^n = \begin{cases}
                1, & \text{if $j\notin \mathcal{U}^{n-1}$},\\
                0, & \text{otherwise}.
                \end{cases}
\end{equation}
Suppose the attention module parameters are denoted as $\bm{\theta}_a = (\mathbf{W}_1, \mathbf{W}_2, \mathbf{v})$, then the weight of pointing to element $j \in \mathcal{N}$ can be calculated as  
\begin{equation}
\tau_j^n = \mathbf{v}^T\tanh(\mathbf{W}_1\mathbf{e}_j + \mathbf{W}_2\mathbf{d}_n), \ \ \  j\in \mathcal{N}, n \in \mathcal{N}.
\end{equation}
The probability of pointing to the index of the $j$-th input sequence can then be found as a softmax function over the weights of the elements not in set $\mathcal{U}^{n-1}$. That is, 
\begin{align}
\begin{split}
\mathbb{P}(u_n = j\ |&\ \tau^n_1, \ldots, \tau^n_{N})= \frac{\exp\big(q_j^n\tau_j^n\big)}{\sum_{i\in\mathcal{N}}\exp\big(q_i^n\tau_i^n\big)}, \quad j, n \in \mathcal{N}. \nonumber
\end{split}
\end{align}
\noindent The parameters of the PtrNet, $\bm{\theta}$, are the concatenation of the encoder, decoder, and the attention module parameters:
\begin{equation}
\bm{\theta} = (\bm{\theta}_e, \bm{\theta}_d, \bm{\theta}_a). 
\end{equation}

{\linespread{1}
\begin{algorithm}[t]
\begin{small}
\caption{REINFORCE-based Parameter Optimization Algorithm}\label{alg:REINFORCE}
\begin{algorithmic}[1]
\Statex \textbf{Input:}  Number of episodes $N_\text{epi}$, time steps per episode $T_\text{epi}$, moving average parameter $\lambda$, learning rate $\epsilon$, and batch size $K_\text{batch}$
\State Initialize $\bm{\theta}$ and the baseline function $b_\text{B}$
\For{$t = \{1, 2, \ldots, T_\text{epi}N_\text{epi}\}$} 
\State System samples $\mathbf{s}_i$, $ i \in \{1,2, \ldots, K_\text{batch}\}$
\State System samples $\mathbf{u}_i$, $ i \in \{1,2, \ldots, K_\text{batch}\}$
\State $b_\text{B} \leftarrow \lambda  b_\text{B} + (1-\lambda)(\frac{1}{K_\text{batch}}\sum_{i =1}^{K_\text{batch}} \Phi(\mathbf{u}_i\ |\ \mathbf{s}_i))$
\State $g_{\bm{\theta}} \leftarrow \frac{1}{K_\text{batch}}\overset{K_\text{batch}}{\underset{i=1}{\sum}}\bigg[\left(\Phi(\mathbf{u}_i\ |\ \mathbf{s}_i) - b_B\right)\nabla_{\bm{\theta}}\log p_{\bm{\theta}}(\mathbf{u}_i\ |\ \mathbf{s}_i)\bigg]$ 
\State $\bm{\theta} \leftarrow \text{ADAM}( \bm{\theta}, g_{\bm{\theta}}, \epsilon)$~\cite{kingma2014adam}
\EndFor
\State \Return $\bm{\theta}^*\leftarrow \bm{\theta}$
\end{algorithmic}
\end{small}
\end{algorithm}
}
\subsection{REINFORCE-based Parameter Optimization Algorithm}
In this subsection, we present an algorithm to obtain the optimal parameters $\bm{\theta}^*$. In the original paper~\cite{Vinyals2015Pointer} that introduced the PtrNet, Vinyals \textit{et al.} proposed to use supervised learning-based methods to train the PtrNet. Later, in \cite{bello2017neural}, Bello \textit{et al.} proposed to use a DRL-based method to train the network parameters. Since using supervised training method requires the optimal solutions of sample problems as training labels, which are computationally intensive to generate, we follow the latter approach.

Since the optimization objective is to maximize the expected aggregate data rate, let us define the performance of $\bm{\theta}$ given $\mathbf{s}$ as $J(\bm{\theta}\ |\ \mathbf{s})$, where
\begin{equation}
J(\bm{\theta}\ |\ \mathbf{s}) =  \mathbb{E}_{\mathbf{u}\sim p_{\bm{\theta}(\cdot|\mathbf{s})}}\left[\Phi(\mathbf{u}\ |\ \mathbf{s})\right].
\label{eq:Reinforce_Performance}
\end{equation}
The gradient of (\ref{eq:Reinforce_Performance}) can be found by the policy gradient theorem~\cite{sutton2017reinforcement} 
\begin{equation*}
\nabla_{\bm{\theta}}J(\bm{\theta}\ |\ \mathbf{s}) = \underset{{\mathbf{u}\sim p_{\bm{\theta}(\cdot|\mathbf{s})}}}{\mathbb{E}}\bigg[\left(\Phi(\mathbf{u}\ |\ \mathbf{s}) - b_B\right)\nabla_{\bm{\theta}}\log p_{\bm{\theta}}(\mathbf{u}\ |\ \mathbf{s})\bigg],
\end{equation*}
\noindent where $b_B$ is any baseline function that reflects the expected aggregate data rate increases as training continues. The reason behind including the baseline function during the training process is to reduce the variance of the gradient. In this paper, we use the moving average of the aggregate data rate as the baseline function. At each training step, we sample a batch of $K_{\text{batch}}$ independent training samples, and estimate the expectation of the gradient via the empirical average. The detailed description of the REINFORCE-based parameter optimization algorithm is shown in Algorithm~\ref{alg:REINFORCE}.

\section{Performance Evaluation}

In our simulations, we consider a network with an area of $100\times100~\text{m}^2$, where the locations of the UEs are sampled from a uniform random distribution within the network area. The transmission power $P$ is set to $1$~W. The noise variance is set to $4\times10^{-9}$~W. The pathloss coefficient $\beta$ is set to 4. In the implementation of PtrNet, between the input sequence and PtrNet, we inserted a fully-connected embedding layer with size 128, to bring the input sequence into a higher dimension. We chose $H = 100$, $K_\text{batch} = 128$, $\lambda =0.9$, $\epsilon = 10^{-3}$, and $T_\text{epi} = 10,000$. Algorithm 1 is implemented using the PyTorch toolbox~\cite{PyTorch}. Simulations are performed on the Compute Canada research computing platform.

%--------------------------------------------------------------------------------
\begin{figure}[t]
    \centering
    \includegraphics[width=0.5\textwidth]{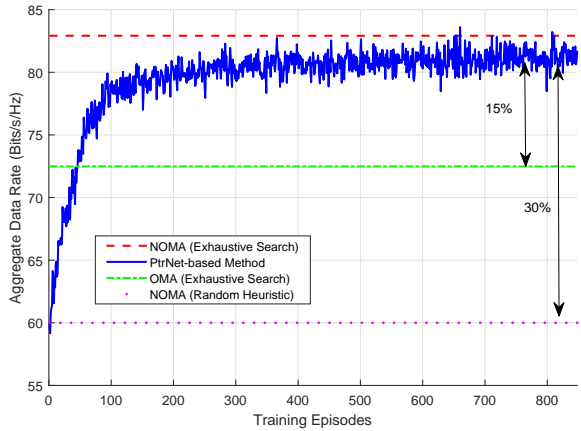}
         \vspace{-0.5cm}
        \caption{Comparison of the aggregate data rate performance of the PtrNet-based solution with three different methods, as the PtrNet parameters are being updated by the REINFORCE algorithm. Five base stations, located at $l_1 = (0, 0)$~m, $l_2 = (25, 25)$~m, $l_3 = (25, -25)$~m, $l_4=(-25, 25)$~m, and $l_5 =(-25, -25)$~m, are deployed to serve ten UEs.}
    \label{fig:plot1}
     \vspace{-0.5cm}
\end{figure}
\begin{figure}[t]
    \centering
    \includegraphics[width=0.5\textwidth]{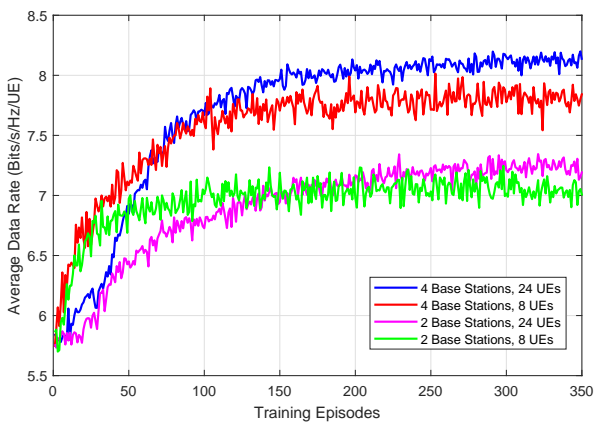}
     \vspace{-0.5cm}
        \caption{The average data rate of the proposed joint user pairing and association algorithm with two and four base stations being deployed, where 8 and 24 UEs are served in each time slot.  For the two base stations case, the base stations are located at $l_2$ and $l_5$. For the four base stations case, the base stations are located at $l_2$, $l_3$, $l_4$, and $l_5$.}
    \label{fig:plot2}
    \vspace{-0.6cm}
\end{figure}

In Fig.~\ref{fig:plot1}, we compare the aggregate data rate performance of the PtrNet-based solution, as the PtrNet parameters are being updated by the REINFORCE algorithm, with the optimal NOMA joint user pairing and association and the random pairing heuristic. The performance of optimal OMA joint user pairing and association is also included to demonstrate the superiority of NOMA compared to OMA. Five base stations are deployed to serve ten UEs. The optimal joint user pairing and association strategies of OMA and NOMA are obtained through exhaustive search, while the random pairing heuristic chooses random permutation of elements in $\mathcal{N}$ as the output sequence $\mathbf{u}$. We observe that the aggregate data rate performance of the proposed PtrNet-based solution begins to stabilize after 200 training episodes, and achieves near-optimal performance with an optimality gap less than 2\%. It can achieve an improvement in aggregate data rate by 15\% compared to the optimal OMA-based approach and by 30\% compared to the random pairing heuristic. In Fig.~\ref{fig:plot2}, we plot the average data rate of the proposed joint user pairing and association algorithm with two and four base stations being deployed and 8 and 24 UEs are being served in each time slot. We observe that having more base stations leads to an improvement in the average data rate, while having more users requires a longer training time for convergence.

\section{Conclusion}
In this paper, we proposed a CSI-based joint user pairing and association scheme for NOMA, based on the PtrNet architecture. We adopted a DRL-based method to find the optimal parameters of the PtrNet, which eliminated the need for the optimal solutions as training labels. Simulation results show that the proposed PtrNet-based solution achieved near-optimal performance. It significantly improved the aggregate data rate of the system, compared to the NOMA random pairing heuristic and the optimal user pairing and association using OMA. Future work will consider the effect of having uncertainties in CSI measurements. We will also investigate the effect of inter-cell interference by allowing different base stations to use the same PRB. 
\vspace{-0.1cm}
\section*{Acknowledgement}
%\vspace{-0.2cm}
This work was supported by Rogers Communications Canada Inc.
\vspace{-0.2cm}
\bibliographystyle{IEEEtran}
\bibliography{IEEEabrv,references}

%---------------------------------------------
\end{document}